\documentclass[10pt]{article}

\usepackage{algorithm}
\usepackage{graphicx}
\usepackage{textcomp}
\usepackage{xcolor}

\usepackage[utf8]{inputenc}
\usepackage{listings}
\usepackage{float}
\usepackage{subfigure}
\usepackage{subcaption}
\usepackage{subfigure}
\usepackage{caption}
\usepackage{caption}

\usepackage{url}
\usepackage{multirow}
\usepackage{cancel}
\usepackage{commath}
\usepackage{algpseudocode}
\usepackage{etoolbox}
\usepackage{varwidth}
\usepackage{tikz}
\usetikzlibrary{shapes,arrows}
\usepackage{colortbl}
\usepackage[margin=1in]{geometry}
\usepackage{hyperref}
\usepackage{cleveref}

\newcolumntype{P}[1]{>{\centering\arraybackslash}m{#1}}

\definecolor{codegreen}{rgb}{0,0.6,0}
\definecolor{codeblack}{rgb}{0.30,0.30,0.30}
\definecolor{codepurple}{rgb}{0.58,0,0.82}
\definecolor{backcolour}{rgb}{0.95,0.95,0.92}
 
\lstdefinestyle{codeListStyle}{
    backgroundcolor=\color{backcolour},
    commentstyle=\color{codegreen},
    keywordstyle=\color{magenta},
    numberstyle=\scriptsize\color{codeblack},
    stringstyle=\color{codepurple},
    breaklines=true,
    breakatwhitespace=true,
    numbers=left,
    xleftmargin=0.2in,
    xrightmargin=0.05in,
    basicstyle=\ttfamily\footnotesize,
    numbersep=5pt,
    frame=single,                  
    tabsize=2
}

\AtBeginDocument{%
  \providecommand\BibTeX{{%
    \normalfont B\kern-0.5em{\scshape i\kern-0.25em b}\kern-0.8em\TeX}}}

\begin{document}

\title{Static Reuse Profile Estimation for Array Applications}

\author{Abdur Razzak$^1$,  Atanu Barai$^2$, Nandakishore Santhi$^2$, Abdel-Hameed A. Badawy$^{1,2}$}

\date{
    \small
    ${}^1$ Klipsch School of ECE, New Mexico State University, Las Cruces, NM 80003, USA\\
    ${}^2$ Los Alamos National Laboratory, Los Alamos, NM 87545, USA\\
    \{arazzak, badawy\}@nmsu.edu, \{abarai, nsanthi\}@lanl.gov\\
}

\maketitle

\begin{abstract}
Reuse distance analysis is a widely recognized method for application characterization that illustrates cache locality. Although there are various techniques to calculate the reuse profile from dynamic memory traces, it is both time and space-consuming due to the requirement to collect dynamic memory traces at runtime. In contrast, static analysis reuse profile estimation is a promisingly faster approach since it is calculated at compile time without running the program or collecting memory traces. This work presents a static analysis technique to estimate the reuse profile of loop-based programs. For an input program, we generate a basic block-level control flow graph and the execution count by analyzing the LLVM IR of the program. We present the memory accesses of the application kernel in a compact bracketed format and use a recursive algorithm to predict the reuse distance histogram. We deploy a separate predictor that unrolls the loop(s) for smaller bounds and generates a temporary reuse distance profile for those small cases. Using these smaller profiles, the reuse profile is extrapolated for the actual loop bound(s). We use this reuse profile to predict the cache hit rate. Results show that our model can predict cache hit rates with an average accuracy of 95\% relative to the dynamic reuse profile methods.
\end{abstract}


\maketitle

\section{Introduction}
\label{sec:intro}
In the current landscape of high-performance computing (HPC), characterized by advances that integrate billions of transistors on a chip, it is crucial for applications to harness the immense computing power available. 
Performance modeling and application profiling play significant roles in helping us understand how to utilize a computing system more efficiently. Reuse distance profiles are a popular metric to capture applications' caches and memory performance. 
It helps to calculate cache hit/miss rates as a metric of program performance. 
These two metrics are essential for memory modeling. They provide valuable insights into application behavior and help optimize performance by identifying potential bottlenecks in accessing memory.

Reuse distance (RD) represents the number of unique memory references between two consecutive references to the same memory address. A reuse profile (RP) is the histogram of the reuse distances for all the memory references of a program. Since RP is platform-independent, it has been an active research topic for calculating reuse distance histograms correctly and efficiently. There is a substantial amount of research proposing both static~\cite{rd_static_narayan, Meng-Ju, dr-chen-ding, matlab, LLVM_Atanu} and dynamic~\cite{PARDA:Niu, l2-cache, ReuseTracker} approaches. In the dynamic approach, dynamic memory traces are collected by running an instrumented program, and then the reuse profile is calculated from the trace.
PARDA~\cite{PARDA:Niu} calculates the reuse profile from a program’s runtime memory trace using a tree-based parallel algorithm. ReuseTracker~\cite{ReuseTracker} is a recent investigation on the parallelization of the dynamic reuse profile calculation approach. Although these tools compute the reuse profile quickly and accurately from a given memory trace, they require the program to be instrumented and executed and traces to be collected before the calculation can begin. This trace collection step is time-consuming, making the process lengthy for bigger problem sizes. On the contrary, static prediction analyzes the compile-time characteristics of a program and estimates the reuse profile and, hence, performance metrics. Since this approach does not require program execution or memory trace collection, it is generally faster than dynamic models. However, without entire trace collection, static techniques lack complete insight into memory reference interleaving, which often leads to issues with prediction accuracy. Researchers have proposed various approaches to calculate reuse profiles statically. However, these methods are limited to specific problem types or have lower accuracy compared to the dynamic tools.

\begin{figure}[b]
    \centering
    \lstset{style=codeListStyle}
    \begin{lstlisting}[language=C,numbers=left]
      int i, j, k, alpha = 99;
      int A[500][500], B[500][500];
    
      for (i = 0; i < 300; i++)
        for (j = 0; j < 200; j++)
          for (k = 0; k < 102; k++){
                A[i][k] = alpha * A[i][k];
                B[j][k] = alpha;
          }
    \end{lstlisting}
    \caption{Example code of nested loops and arrays}
    \label{fig:example_program}
\end{figure}

\begin{figure}[tb]
    \centering
    \includegraphics[trim=0mm -5mm 0mm 0mm, clip, width=.35\textwidth]{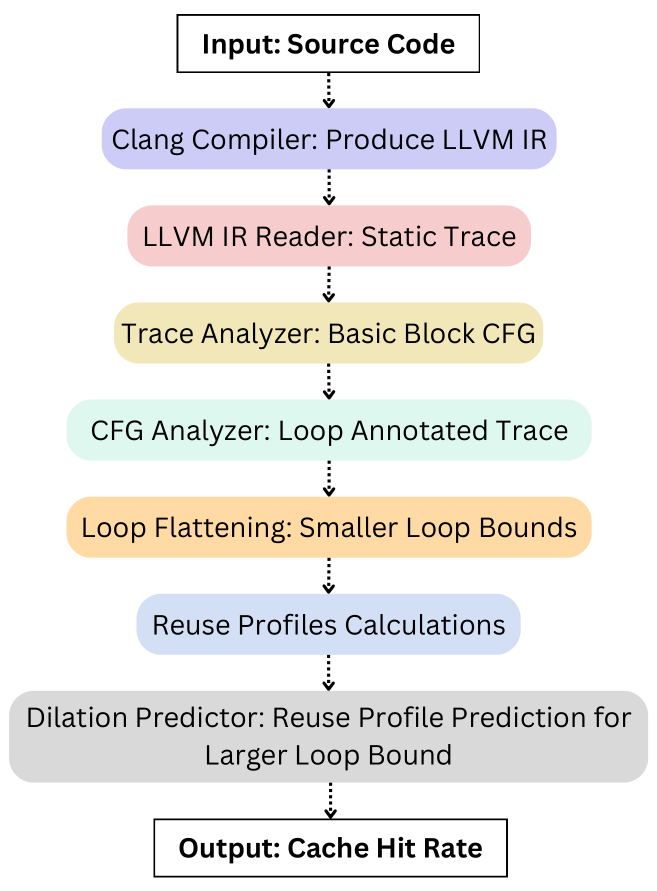}
    \vspace{-5mm}
    \caption{Steps of static analysis based reuse profile prediction}
    \vspace{-5mm}
    \label{fig:overview}
\end{figure}

This paper introduces a static approach to estimating reuse profiles. We first utilize the LLVM~\cite{Lattner:LLVM} tool to generate the given program's intermediate representation (IR). It helps to produce the program's closed-form representation in a bracketed format using ~\cite{LLVM_Atanu}. We generate a smaller problem by replacing the loop bounds with smaller numbers. Then, we perform loop unrolling for those smaller problems and calculate the reuse profiles for a few different smaller loop bounds of the program. By analyzing smaller loop-bound reuse profiles of the same program, we establish some mathematical relationships between the reuse distance and the loop bounds and extrapolate the reuse profiles for the actual given program. Finally, this reuse profile is used to predict the cache hit rate of the program given the cache configurations. Figure~\ref{fig:overview} provides a high-level overview of the steps involved in the proposed approach.

We compare the reuse distance and cache hit rates of the applications from Table~\ref{tab:test_applications} with the existing state-of-the-art and widely used dynamic tool PARDA~\cite{PARDA:Niu}. The results demonstrate that our static model predicts cache hit rates with an average of 95\% accuracy compared to PARDA while achieving a significant speedup in calculation time. We believe modeling program characteristics such as reuse distance and cache hit rate may involve various strategies, however, the accuracy and speed of our static model underscore its potential as a valuable contribution to this domain.

The rest of the paper is structured as follows: Section ~\ref{sec:related_works} provides an overview of previous and contemporary works on both dynamic and static methods. The detailed mechanics of the proposed static model are explained step by step in Section ~\ref{sec:method}. We introduce the reference metrics and result comparison with the dynamic tool in Section ~\ref{sec:results}. Section ~\ref{sec:limitations} and ~\ref{sec:conclusion} show the limitations of this research and conclusion respectively.

\section{Related Work}
\label{sec:related_works}
Researchers proposed several approaches to calculate reuse profiles of program kernels efficiently. Among the static techniques, Narayanan~\emph{et al.}~\cite{rd_static_narayan} attempted to calculate the reuse profile using a syntax tree. However, this technique does not cover the array inside a loop scenario. Liu~\emph{et al.} introduced Parallel Locality Using Static Sampling (PLUSS)~\cite{dr-chen-ding} to predict cache miss ratios with OpenMP-parallelized method using symbolic models: the Negative Binomial Distribution (NBD) for cache sharing and the Racetrack model for data sharing. This approach efficiently analyzes cache performance across thread configurations. However, they are limited to the assumption of regular application codes without dynamic memory access patterns and uniform thread speeds with balanced loads. Wu~\emph{et al.}~\cite{Meng-Ju} investigated reuse distance on multicore cache hierarchy scalability. They also showed that interference-based locality degradation is more significant than sharing-based locality degradation. However, their model is limited in accuracy for applications with unpredictable memory access patterns. Chauhan~\emph{et al.}~\cite{matlab} presented a static algorithm to calculate reuse distance from MATLAB's high-level characteristics at the source code level. However, they did not demonstrate how their approach handles scenarios where arrays are used inside a loop that enables memory references to change unpredictably. Additionally, the method's accuracy decreases with non-uniform references, making it more computationally expensive. Ding~\emph{et al.}~\cite{pattern-recognition}, on the other hand, presented an approximate analysis of reuse using the Least Recently Used (LRU) stack distance, followed by applying a pattern recognition algorithm to the distance-based sampling. Nevertheless, the changing references of an array inside a loop do not follow the consistent distance pattern as they described because of arbitrary use of the array references if the loop's variable controls them, and those cases are not taken care of in their study. Likewise, the method did not find a generalized pattern-making solution for other programs. Recently, Barai~\emph{et al.}~\cite{LLVM_Atanu} has proposed a reuse distance calculation method using LLVM-based techniques, but it cannot calculate reuse profiles for applications with arrays within loops.

Significant research has also been conducted on the dynamic approach. Niu~\emph{et. al.}, presented PARDA~\cite{PARDA:Niu} which  parallelizes reuse distance calculation. They partitioned the runtime memory traces into \emph{n} chunks, then processed them in parallel by n processors. A tree-based algorithm was employed to calculate the reuse distance. Due to the optimization, they achieved a speedup of 13 -- 50x  compared to the serial calculation. However, collecting dynamic traces is mandatory, which is a time-intensive process. Keramidas~\emph{et. al.}~\cite{l2-cache} revealed a novel method for cache replacement based on dynamic reuse-distance prediction, which directly uses instruction-based predictions to optimize L2 cache replacement policies. However, their method does not perform well in scenarios with highly irregular or unpredictable memory access patterns.

Our method employs a static approach that leverages LLVM IR to predict reuse profiles and cache hit rates, which are crucial performance metrics. This strategy eliminates the need for runtime memory trace collection, resulting in substantial time savings. The model is validated using six benchmark programs, and the results are remarkably similar to those of contemporary dynamic models.

\section{Methodology}
\label{sec:method}
This section comprehensively explains each step in our proposed methodology, with the workflow illustrated in Figure~\ref{fig:overview}. Our tool begins by taking a source code program as input and then processes it through several steps. Lastly, it generates the program’s reuse profile and cache hit rate. We describe each step in detail using the example code shown in Figure~\ref{fig:example_program} to illustrate the process.

\subsection{LLVM IR using Clang}
First, we generate the source program’s LLVM Intermediate Representation (IR) using the Clang compiler with the command \textbf{clang -g -c -emit-llvm source.c}. Clang performs pre-processing and compilation to generate LLVM IR, and the output is in a binary format called bitcode. This bitcode is then used in further optimization and linking steps to produce the final executable. However, using the \textbf{-emit-llvm}, Clang generates the IR in a human-readable format for the \emph{source.c} file while also including debug information.

\subsection{Static Trace from LLVM IR Reader}
\label{SubSubSec:StaticTraceLLVM-IR}
The generated IR file contains detailed basic block information, which is crucial for understanding the program’s control flow. A basic block is a linear sequence of instructions without branches, with one entry and one exit point. The IR reader then processes this file, translating the human-readable IR code into an internal data structure that encapsulates all necessary program details, including basic blocks, control flow, and data dependencies. The static traces along with the basic block information derived from this process are essential for accurately modeling program behavior.

\subsection{Trace Analyzer}
\label{subsubsec:trace_analyzer}
There are two sub-parts of the Trace analyzer.

\subsubsection{Producing Control Flow Graph}
In this step, we generate the program's control flow graph (CFG). Each node in the CFG represents the program's basic blocks. This graph shows all successors and predecessors. Vertices are directional and point to source and destination nodes. The vertices are weighted, which represents the probability of going that path. This probability is calculated based on the program's input set.

\subsubsection{Basic Block Execution Counts}

The calculation of basic block execution counts involves analyzing the program’s Control Flow Graph (CFG) along with branch probabilities, which indicate the likelihood of each branch being taken during execution. By leveraging these branch probabilities, we determine the exact number of times each basic block is executed. These counts are directly connected to the program’s inputs, as different inputs cause the program’s basic blocks to be executed a different number of times. If the input size increases, some basic block execution frequencies also increase. Each memory reference within a basic block is executed the same number of times the block is executed. As a result, determining the execution counts of basic blocks helps to predict the frequency of specific memory references, which is crucial for subsequent analyses in this study.

\subsection{Loop Annotated Trace from CFG Analyzer}
\label{subsubsec:cfg_analyzer}
We now have the CFG, branch probabilities, and basic block execution counts. Using all this information, we determine the program execution path with maximum probability within the CFG. By analyzing that path, we determine any program loop or circular connection. Then, we mark the boundary of the loop starting position and the end to mark it as a loop-bound area. Thus, we have a loop-annotated static memory trace of the program.

Next, we identify the variables used to index the arrays. We track the LLVM GEP instructions in the IR file to identify the index variables and array references. Once identified, these variables are annotated in the static trace. Usually, the array index variable is loaded from memory just before the array is accessed. For example, in the trace, arrayidx8 is a two-dimensional array in the program. Therefore, the two previously accessed IR variables are this array’s loop-controlling variables. Finally, we determine the variable that controls the loop and annotate the static trace with this information. This variable is also accessed just before the loop starts.

In the trace of our example program, arrayidx8 is a two-dimensional array; therefore, the two previously accessed variables are this array’s loop-controlling variable (arrayidx8[i][k]). Likewise, the reference ‘i’ is accessed just before the first loop bound starts, indicating that i is the loop-controlling variable. 

After this step, the trace looks like the following.

\noindent
\textsf{\textbf{retval, alpha, i, [300\textasciitilde i, i, j, [200\textasciitilde j, j, k, [102\textasciitilde k, k, alpha, i, k, A\textasciitilde i\textasciitilde k, i, k, A\textasciitilde i\textasciitilde k, alpha, j, k, B\textasciitilde j \textasciitilde k, k, k ], k, j, j ], j, i, i], i}}

\subsection{Loop Flattening: Smaller Loop Bounds}
Our final goal is to calculate the reuse distance histogram and cache hit rate for larger loop bounds. Therefore, we need to analyze the behavior of the array’s variable memory references with some smaller loop bounds. To do that, the reuse profile is initially calculated for smaller loop bounds, allowing us to observe the behavior of changes in the reuse profiles if the loop bound increases by a small number. We can prepare a change factor by analyzing these changes, which is then applied to larger loop bounds. To accurately calculate reuse profiles for these smaller loop bounds, it is essential to flatten the loop-annotated memory traces. This process provides the necessary data for making informed predictions about larger loop bounds.

\begin{figure}[tb]
    \centering
    \includegraphics[trim=0mm -5mm 0mm 0mm, clip, width=.35\textwidth]{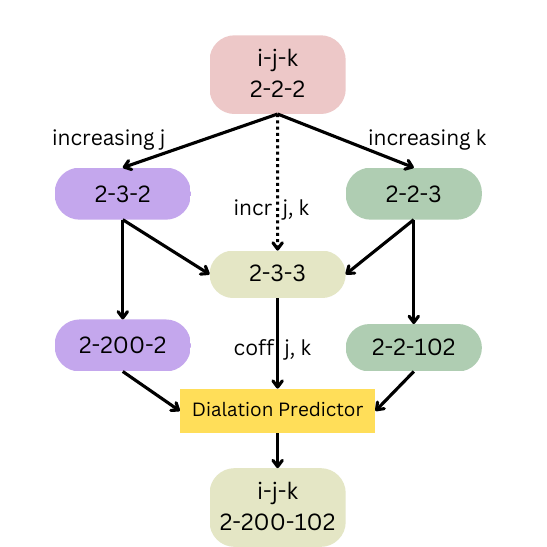}
    \caption{Loop flattening for predicting reuse profile for higher loop bound from smaller bounds.}
    \label{fig:loop_flatten}
\end{figure}

Figure~\ref{fig:loop_flatten} shows the process of making 2-2-2 loop bounds as our baseline and flattening the whole loop. After the loops are flattened, the trace looks as follows.

\textbf{\newline
['i', 'j', \newline
‘j’, ‘k’, \newline
‘k’, ‘alpha’, ‘i’, ‘k’, ‘A\textasciitilde i\textasciitilde k-0-0’, ‘i’, ‘k’, ‘A\textasciitilde i\textasciitilde k-0-0’, ‘alpha’, ‘j’, ‘k’, ‘B\textasciitilde j\textasciitilde k-0-0’, ‘k’, ‘k’,\newline
‘k’, ‘alpha’, ‘i’, ‘k’, ‘A\textasciitilde i\textasciitilde k-0-1’, ‘i’, ‘k’, ‘A\textasciitilde j\textasciitilde k-0-1’, ‘alpha’, ‘j’, ‘k’, ‘B\textasciitilde j\textasciitilde k-0-1’, ‘k’, ‘k’,\newline
‘k’, ‘j’, ‘j’, \newline
‘j’, ‘k’, \newline
‘k’, ‘alpha’, ‘i’, ‘k’, ‘A\textasciitilde i\textasciitilde k-0-0’, ‘i’, ‘k’, ‘A\textasciitilde i\textasciitilde k-0-0’, ‘alpha’, ‘j’, ‘k’, ‘B\textasciitilde j\textasciitilde k-1-0’, ‘k’, ‘k’,\newline
‘k’, ‘alpha’, ‘i’, ‘k’, ‘A\textasciitilde i\textasciitilde k-0-1’, ‘i’, ‘k’, ‘A\textasciitilde j\textasciitilde k-0-1’, ‘alpha’, ‘j’, ‘k’, ‘B\textasciitilde j\textasciitilde k-1-1’, ‘k’, ‘k’,\newline
‘k’, ‘j’, ‘j’,\newline
'j', 'i', 'i', ......]
}

Now, we generate flattened traces for loop bounds 2-2-3 to observe how they change in the reuse profile and mark the impact of increasing k by 1. In the same way, from analyzing the 2-3-2 profile, we observe the increase of 1 j. After that, we observe the reuse profile for loop bounds 2-3-3 where both j and k increased by 1. Here, we try to find a coefficient for j and k increases. After building a connection from these small bounds, we want to reach 2-200-102 bound. This process is explained in the next section.

\subsection{Reuse Profile Calculation}
\label{subsec:ReuseProfileCalc}
We calculate the reuse profile for the flattened stack traces individually. We use the least recently used (LRU) stack to keep track of the most recently used addresses. If the item is not found in the LRU stack, it is a cold miss or the first time accessing the address. We put that address into the LRU stack with the index and mark the reuse distance as -1. Otherwise, if it is found on the LRU stack, we take a hash set between the current index and the LRU stack stored index. That filters out the duplicate references in the middle. The size of the unique set is the reuse distance of that address. We store the count of unique addresses in a dictionary. That dictionary turns into a reuse profile at the end. The algorithm~\ref{alg:calc_reuse_profile_flattened} shows the calculation method.

\begin{algorithm}[tbp]
\begin{algorithmic}[1]    \Procedure{CalcReuseProfileFlattened}{$stack\_trace$}
        \State $rf \gets \{\}$ 
        \State $inf \gets [\,]$
        \State $LRU\_dict \gets \{\}$
        \For{$index, item$ \textbf{in} \textbf{enumerate}($stack\_trace$)}
            \If{$item$ \textbf{not in} $LRU\_dict$} \Comment{If the item not found}
                \State $inf$.append($item$)
                \State $LRU\_dict[item] \gets index$
            \Else
                \State $sub\_list \gets stack\_trace[LRU\_dict[item]+1 : index]$ \Comment{Take all subset between two refs}
                \State $unique\_items \gets$ \textbf{list(set}($sub\_list$)) \Comment{Remove duplicates}
                \State $reuse\_distance \gets$ \textbf{len}($unique\_items$)
                \If{$reuse\_distance$ \textbf{in} $rf$}
                    \State $rf[reuse\_distance] \gets rf[reuse\_distance] + 1$
                \Else
                    \State $rf[reuse\_distance] \gets 1$
                \EndIf
                \State $LRU\_dict[item] \gets index$ \Comment{Updating with the least recently used index}
            \EndIf
        \EndFor
        \State \Return $rf, inf$
    \EndProcedure
\end{algorithmic}
\caption{Calculating Reuse Profile}
\label{alg:calc_reuse_profile_flattened}
\end{algorithm}

\subsection{Dilation Prediction}
Following the previous steps, we can calculate reuse profiles for small basic loop bounds. However, from these smaller reuse profiles, we need to calculate the reuse profile for larger loop bounds as shown in the example program in Figure~\ref{fig:example_program}. To achieve this target, we first identify the changes in the reuse profile due to scaling in the inner loop bound and then we address the impact of incorporating the outer loops.

\subsubsection{Reuse Profile Prediction: Scaling Inner Loop Bounds}

In this step, we focus on the innermost loop, where the array references change only with the innermost loop index variable while keeping other array positions constant. We predict the inner loop first because it repeats with each iteration of the outer loop. By accurately predicting the inner loop's reuse profile, we can determine the outer bounds by analyzing its repeated behavior later.

The figure~\ref{fig:one_loop_example} illustrates a code snippet featuring two-dimensional arrays, A and B, used within a loop. The first dimension remains constant, while the second dimension depends on the loop variable k, which iterates from 0 to 101. This setup demonstrates the impact of a single loop variable altering the array reference locations.

\begin{figure}[tb]
    \centering
    \lstset{style=codeListStyle}
    \begin{lstlisting}[language=C,numbers=left]
    for(k=0; k<102; k++) {
        A[0][k] = alpha * A[0][k];
        B[0][k] = alpha;
    }
    \end{lstlisting}
    \vspace{-3mm}
    \caption{Array references are changing a single loop.}
    \vspace{-3mm}
    \label{fig:one_loop_example}
\end{figure}

Now, if we produce a few loop annotated traces and reuse profiles for the smaller bounds starting from the value of 2, we can create a relation between the numbers of each reuse distance. The loop annotated trace for Figure~\ref{fig:one_loop_example} looks like the following if $k=2$.
\textbf{\newline
[retval, k, [2, k, alpha, k, A\textasciitilde0\textasciitilde k, k, A\textasciitilde0\textasciitilde k, alpha, k, B\textasciitilde0\textasciitilde k, k, k, ], k]]
}

Similarly, we compute the static trace from loop unrolling for $k=3$, and $k=4$. Using the reuse profile calculation technique explained in Section~\ref{subsec:ReuseProfileCalc}, we find the reuse profiles shown in Table~\ref{tab:one_loop_rf}.

\begin{table}[tb]
    \centering
    \begin{tabular}{|c|c|}
        \hline
        \textbf{K (Inner Loop Bound)} & \textbf{Reuse Distance Histogram} \\ \hline
        2 & \{0: 5, 1: 9, 2: 5,  -1: 7\} \\ \hline
        3 & \{0: 7, 1: 13, 2: 8,  -1: 9\} \\ \hline
        4 & \{0: 9, 1: 17, 2: 11,  -1: 11\} \\ \hline
    \end{tabular}
    \caption{Reuse profiles for smaller inner loop bounds}
    \label{tab:one_loop_rf}
\end{table}

We observe the number sequence from these small cases and find a linear connection between the bound k and reuse distance frequency increment. Considering the values in $k=2$ as a baseline, we can create the following linear equation.

\begin{equation}
\label{eq:one_loop_rule}
    Frq = B_{2}  + Dist_K \times Incr_K  
\end{equation}

Using this equation, we predict the frequency of each reuse distance where considering $B_{2}$ is as the base frequency at $k=2$, $Dist_K$ is the loop bound distance that we want to reach, \textit{e.g.}, from 2 to 102 $Dist_K$ is 100 according to the example code in Figure~\ref{fig:one_loop_example}, and the $Incr_K$ represents the changes in the frequency for each increment of k, such as $k=2$ to $k=3$, the reuse distance for 0 increases by 2, 4 increases for reuse distance 1, 3 for 2 and 2 for -1. Table~\ref{tab:one_loop_pred} shows the frequency prediction for each reuse distance for $k=102$ by following equation~\ref{eq:one_loop_rule}, and it exactly matches the dynamic reuse profile. 

\begin{table}[bt]
    \centering
    \begin{tabular}{|c|c|c|}
        \hline
        \textbf{RD} & \textbf{Equation} & \textbf{Pred. Frequency} \\ \hline
        0 & 5 + 100 * 2 & 205 \\ \hline
        1 & 9 + 100 * 4 & 409 \\ \hline
        2 & 5 + 100 * 3 & 305 \\ \hline
        -1 & 7 + 100 * 2 & 207 \\ \hline
    \end{tabular}
    \caption{Reuse profiles prediction from the linear equation.}
    \label{tab:one_loop_pred}
\end{table}

\subsubsection{Incorporating Outer Loops and Dilation}
In this step, we incorporate the inner loop prediction into outer loops and observe the changes. If the outer loop bound starts incrementing from 2, it shows some changes in the reuse distance number itself, which is not seen in the earlier steps. The loop annotated trace for the program in Figure~\ref{fig:example_program} is shown below, where the loop bounds are set to two.

\textbf{
[‘retval’, ‘alpha’, ‘i’, ‘[2’, ‘i’, ‘j’, ‘[2’, ‘j’, ‘k’, ‘[2’, ‘k’, ‘alpha’, ‘i’, ‘k’, ‘A\textasciitilde i\textasciitilde k’, ‘i’, ‘k’, ‘A\textasciitilde i\textasciitilde k’, ‘alpha’, ‘j’, ‘k’, ‘B\textasciitilde j\textasciitilde k’, ‘k’, ‘k’,’]’, ‘k’, ‘j’, ‘j’,’]’, ‘j’, ‘i’, ‘i’,’]’, ‘i’]
}

If k is changed from 2 to 3 and 4, we notice that the frequency of some reuse distances also changes, such as for the distances 0,1,2,3,4,5 and -1, which is unchanged in every reuse profile. On the other hand, some reuse distances are newly introduced or discarded from the reuse distance list. For example, reuse distance 7 is present in the $k=2$ loop bound; however, it is missing in $k=3$ or $k=4$.

\begin{table}[tb]
    \centering
    \begin{tabular}{|c|p{7cm}|}
        \hline
        \textbf{K} & \textbf{Reuse Distance Histogram} \\ \hline
        2 & \{0: 35, 1: 11, 2: 37, 3: 25, 4: 5, 5: 12, \textbf{\textit{7: 4, 9: 1, 10: 2, 11: 1}}, -1: 13\} \\ \hline
        3 & \{0: 43, 1: 15, 2: 53, 3: 37, 4: 5, 5: 20, \textbf{\textit{9: 6, 12: 1, 13: 2, 14: 2, 15: 1}}, -1: 17\} \\ \hline
        4 & \{0: 51, 1: 19, 2: 69, 3: 49, 4: 5, 5: 28, \textbf{\textit{11: 8, 15: 1, 16: 2, 17: 2, 18: 2, 19:1}}, -1: 21\} \\ \hline
    \end{tabular}
    \caption{Outer loops shows dilation in reuse distance itself}
    \label{tab:three-loop}
\end{table}

To identify this change, we divide the problem into two distinct parts. The first part involves constant components where the reuse distance numbers remain fixed in every reuse profile while only the frequencies change. These can be predicted using the linear equation technique explained in the previous section. 

The second part involves scenarios where both the reuse distance and frequency fluctuate. For this part, we develop a separate tool to predict the varying reuse distances and frequencies. The predictor utilizes three lists of reuse distances to analyze the change in reuse distances. For example, as in Table~\ref{tab:three-loop}, the three lists are \{7, 9, 10, 11\}, \{9, 12, 13, 14, 15\}, and \{11, 15, 16, 17, 18, 19\}. The predictor examines the relationship between each list, its neighboring lists, and the size dependence on the value of \(k\).

\begin{figure}[tb]
    \centering
    \includegraphics[trim=0mm -5mm 0mm 0mm, clip, width=.35\textwidth]{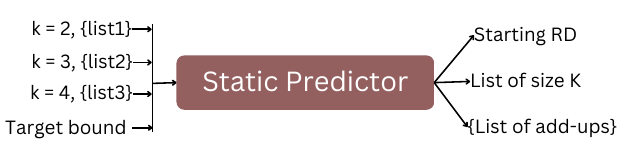}
    \vspace{-5mm}
    \caption{Static predictor for target loop bound from given lists}
    \label{fig:static_predictor}
\end{figure}

From these three lists and target bound, the predictor constructs equations based on the given lists and finally generates three types of output: the starting reuse distance for the changing list when \(k = n\), the size of the list, and the incremental numbers on each position. Figure~\ref{fig:static_predictor} shows the input and output for the static predictor. 

For example, with the given reuse distance lists and a value of \(k = 102\), the predictor outputs the following: starting number - 207, list size - 104, and a cumulative list starting from the initial number, which is \{0, 102, 103, 104, \ldots, 204\}. Using these three outputs, we can now predict the reuse distance list for $k=102$. Similarly, we can calculate the volatile frequency numbers for the 102 k loop bound.

\subsubsection{Estimating for the Outer Loop} 

At this stage, we aim to estimate the reuse profile for the complete bounds of the nested loops with indices \(j\) and \(k\) while keeping the outermost loop with index \(i\) fixed at a bound of 2.  Up to this point, we compute the reuse profiles for various loop configurations, specifically 2-2-2, 2-2-3, 2-3-2, and 2-3-3. By examining these reuse profiles, we can assess the combined impact of incrementing both \(j\) and \(k\) by 1, a relationship we define this coefficient as \(Coff_{JK}\). Figure~\ref{fig:loop_flatten} illustrates the overall approach to this task with a visual representation of the changes in reuse profiles across these smaller loop-bound configurations.

Next, we determine the reuse profile for the innermost loop when its bounds are set to the wanted values, utilizing the technique outlined in the previous section. In addition, we independently analyze the effect of extending the bounds of \(j\)  while keeping the other variable  $k$ constant. This step is crucial in understanding how changes in \(j\) influence the overall reuse profile when \(j\) reaches its full bounds.

From the collective analysis of these profiles, we apply Equation~\ref{eq:balance} to compute the reuse distance frequency. This allows us to quantify how the reuse patterns evolve as the loop bounds are modified. At the end of this step, we achieve the full bounds for both \(j\) and \(k\) while maintaining \(i\) at 2.

\begin{equation}
\label{eq:balance}
    Frq= B_{22}  + Dist_J \times Incr_J + Dist_K \times Incr_K + Coff_{JK} \times Dist_J \times Dist_K 
\end{equation}

In the  equation~\ref{eq:balance}, the frequency of each reuse distance is calculated from $B_{22}$ which is the baseline reuse distance, mainly 2-2-2 is considered the baseline reuse profile; $Dist_J$ is the distance in loop bound that $J$ is going up; $Incr_J$ is the change in frequency for each $J$ increase. Also, it is the same for the terms for the $k$ loop.  Finally, $Coff_{JK}$ represents the coefficient impact of scaling both $J$ and $K$ bounds together.

For loops with three nesting levels, we categorize the problem as a third-degree, three-variable problem since the loop depth is three and the loop variables are 
$i$, $j$, $k$. Equation~\ref{eq:equation_3_degree} illustrates the equation used to model these scenarios.

\begin{align}
\label{eq:equation_3_degree}
    Frq &= B_{222} + Dist_I \times Incr_I + Dist_J \times Incr_J + Dist_K \times Incr_K \notag \\
    &\quad + Coff_{IJ} \times Incr_I \times Incr_J
    + Coff_{JK} \times Incr_J \times Incr_K \notag \\
    &\quad + Coff_{IK} \times Incr_I \times Incr_K  + Coff_{IJK} \times Incr_I \times Incr_J \times Incr_K
\end{align}

Following these rules, we calculate the reuse distance profile for the larger loop bounds, where $i = 300$, $j = 200$, and $k = 102$ in this example. We use this predicted reuse profile to calculate cache hit rates.

\section{Results}
\label{sec:results}
This section presents the experimental results and how our methodology compares to dynamic reuse distance tools. Our investigation into developing a static method for predicting reuse profiles and subsequently calculating cache hit rates has shown promising results, closely matching dynamic reuse profile calculations. The following sections describe our results and analysis. 

\begin{table}[tb]
    \centering
    \resizebox{\linewidth}{!}{%
    \begin{tabular}{||c||P{5.0cm}||c||}
    \hline
    \textbf{App} & \textbf{Domain} & \textbf{Loops and Bounds (outer to inner)}\\
    \hline\hline
    atax & Linear Algebra (LA) & 2 nested; 150-70\\
    \hline
    mvt & LA & 2 nested; 100-200\\
    \hline
    gemver & Basic Linear Algebra Subprograms & 2 nested; 400-500\\
    \hline
    trmm & LA & 2 nested; 1000-1000\\
    \hline
    2mm & LA & 3 nested; 300-200-102\\
    \hline
    3mm & LA & 3 nested; 30-40-50\\
    \hline
    \end{tabular}
    }
    \caption{List of applications used to validate results.}
    \label{tab:test_applications}
\end{table}

\subsection{Comparison Methods \&  Metrics}
\subsubsection{{Cache}}
\label{subsubsec:cache-ref}
We maintain a consistent cache configuration throughout the testing process and report the resulting cache hit rates to ensure a fair comparison between our static predictor and the dynamic calculator. We employ a 64KB cache, 32-way set associative, with a 32B cache line size.

\subsubsection{Dynamic Model}

To evaluate our proposed ideas, we compare our results against the widely accepted tree-based dynamic reuse distance tool, PARDA~\cite{PARDA:Niu}, and assess the accuracy of our approach. PARDA is a parallel reuse profile tool that performs reuse distance analysis 13-50x faster than other dynamic methods. It accurately computes the reuse distance histogram from a given memory trace. However, this tool requires runtime traces of the application to initiate the calculation process. Obtaining these traces requires a significant amount of time. This experiment compares our static method with PARDA to evaluate the accuracy and the time.

\subsubsection{{Applications Used for Validation}}

We use snippets of the nested loops of six programs from the PolyBench~\cite{polybench-github-codes} benchmark suite. PolyBench is a widely used benchmark suit. It consists of various computational kernels from a wide range of domains, such as linear algebra, stencils, data mining, etc. In this experiment, we utilize the applications listed in Table~\ref{tab:test_applications}. We compare reuse profiles and cache hit rates of these selected applications. We use nested loop snippets of these application kernels. The nested loops of the testing programs are evaluated with diverse problem sizes by adjusting different loop bounds. For instance, an application called `2mm` contains three nested loops. We set the outermost loop bound to 300, the middle loop to 200, and the innermost loop to 102. The third column of Table~\ref{tab:test_applications} shows the number of nested loops and their bounds from the outer to the innermost loop.

\subsubsection{{Cache Hit Rate Measurement}}

We utilize an analytical memory model, a stack distance-based cache model (SDCM)~\cite{brehob:analytical} to measure cache hit rates. This tool takes the program's reuse profile and cache parameters and generates the cache hit rates. If we keep the cache parameters fixed while calculating the reuse profile dynamically and statically, variations in the reuse profile will lead to differences in the cache hit rates.

SDCM uses the following equation~\ref{eq:phit} to calculate the probability of a hit {\em P(h)} for the entire reuse profile.

\begin{equation}
\label{eq:phit}
P(h) = \sum\limits_{i=0}^N P({D_i}) \times P({h\mid D_i})
\end{equation}

\noindent where, $P({D_i})$ is the probability of $i^{th}$ reuse distance ($D$) in a reuse distribution and ($P(h\mid D)$) is the conditional probability of hit given a distance $D$. $P({D_i})$ is calculated from the frequency of $D$ in the reuse histogram. These hit rates can be further used in runtime prediction of the applications~\cite{YehiaSc21, ppt-multicore, arafa-ppt-gpu}, which is beyond the scope of this paper.

\subsection{Analysis}
We comprehensively analyze the results by dividing them into three key areas. We use the programs in Table~\ref{tab:test_applications} for testing.

\subsubsection{{Reuse Distance Histogram}}

\begin{figure*}[htbp]
    \centering
    \begin{minipage}{0.49\textwidth}
        \centering
        \includegraphics[width=\textwidth]{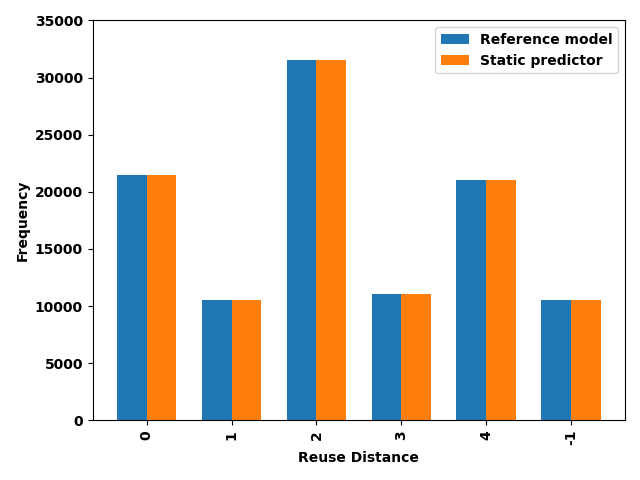}
        \vspace{-6mm} 
        \captionof{subfigure}{Atax two nested loop with 150-70 bounds.}
        \label{fig:atax}
        \vspace{6mm} 
        
    \end{minipage}
    \hfill
    \begin{minipage}{0.49\textwidth}
        \centering
        \includegraphics[width=\textwidth]{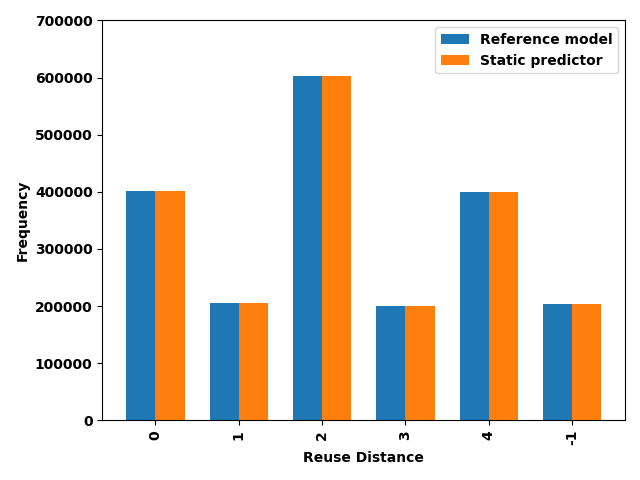}
        \vspace{-6mm} 
        \captionof{subfigure}{Gemver two nested loop with 400-500 bounds.}
        \label{fig:gemver}
        \vspace{6mm} 
    \end{minipage}

    \begin{minipage}{0.49\textwidth}
        \centering
        \includegraphics[width=\textwidth]{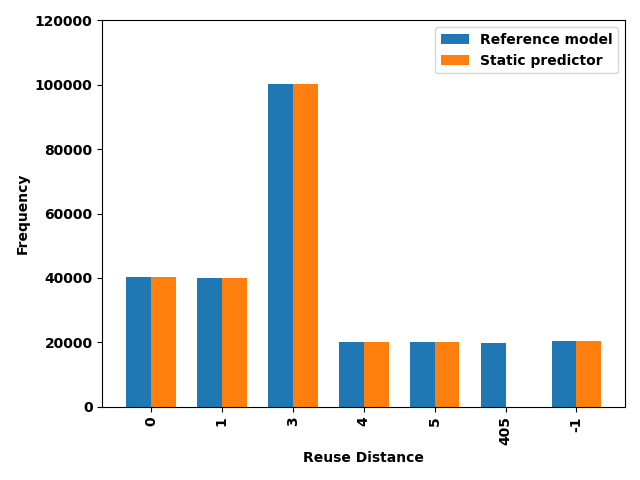}
        \vspace{-6mm} 
        \captionof{subfigure}{Mvt two nested loop snippet with 100-200 bounds.}
        \label{fig:mvt}
        \vspace{6mm} 
    \end{minipage}
    \hfill
    \begin{minipage}{0.49\textwidth}
        \centering
        \includegraphics[width=\textwidth]{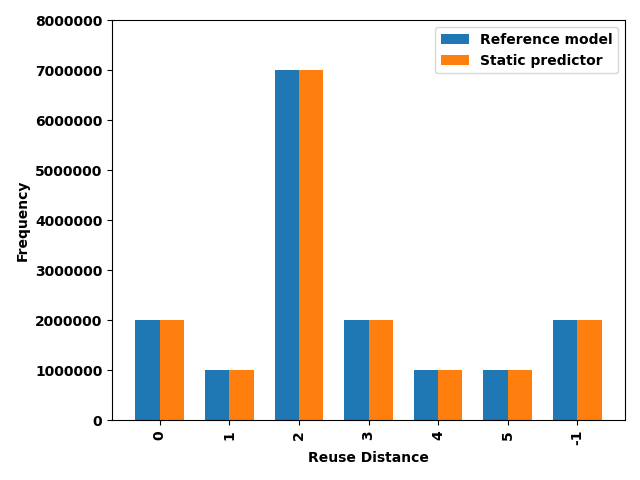}
        \vspace{-6mm} 
        \captionof{subfigure}{Trmm two nested loop with 1000-1000 bounds.}
        \label{fig:trmm}
        \vspace{6mm} 
    \end{minipage}

    \begin{minipage}{0.49\textwidth}
        \centering
        \includegraphics[width=\textwidth]{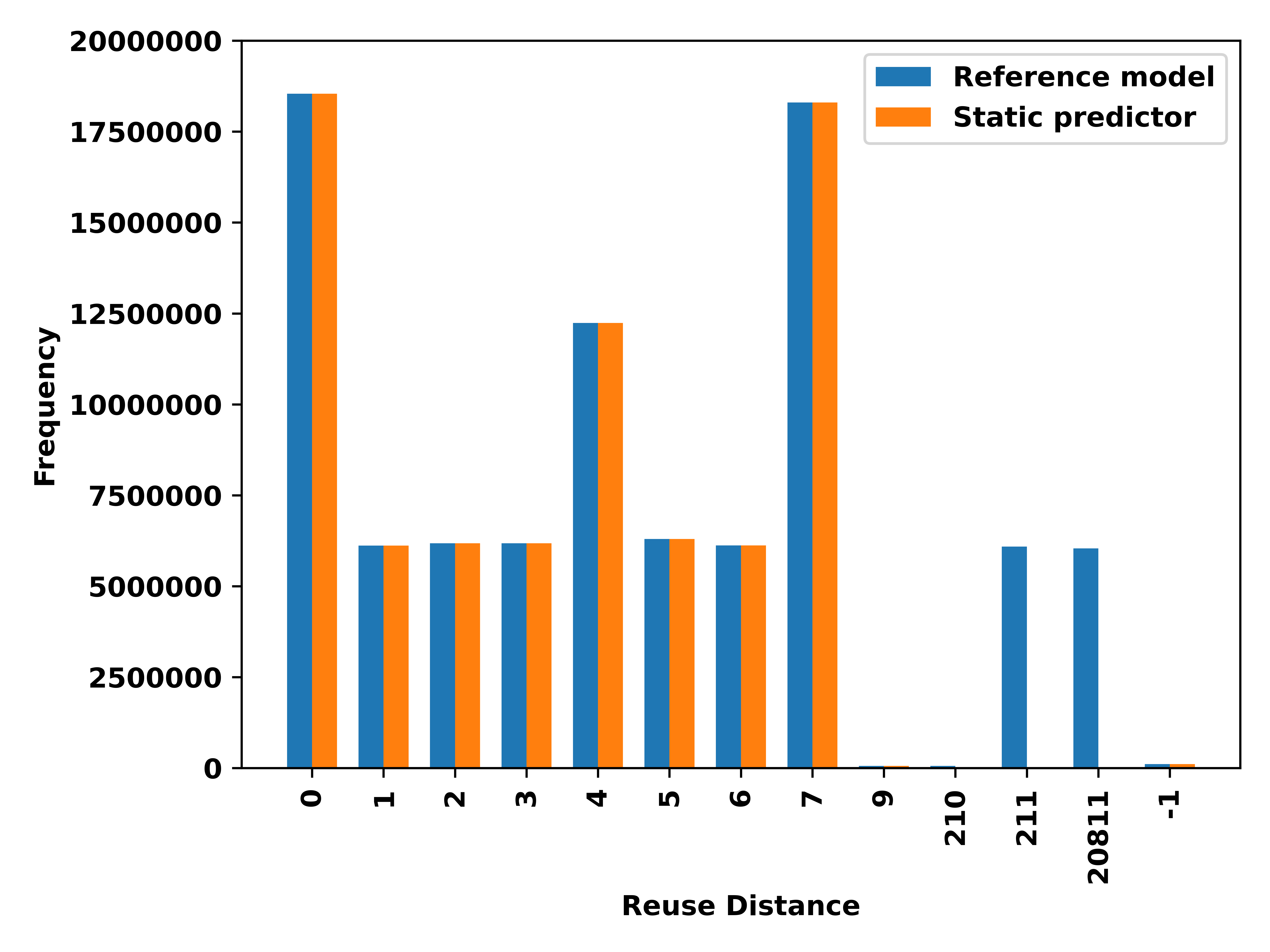}
        \vspace{-6mm} 
        \captionof{subfigure}{2mm three nested loop with 300-200-102 bounds.}
        \label{fig:two-mm}
        \vspace{6mm} 
    \end{minipage}
    \hfill
    \begin{minipage}{0.49\textwidth}
        \centering
        \includegraphics[width=\textwidth]{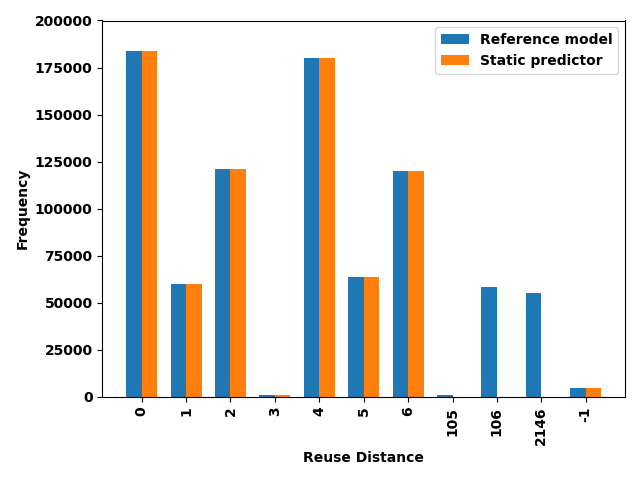}
        \vspace{-6mm} 
        \captionof{subfigure}{3mm three nested loop with 30-40-50 bounds.}
        \label{fig:three-mm}
        \vspace{6mm} 
    \end{minipage}
    \vspace{-6mm} 
    \caption{Reuse Distance comparison between the dynamic tool \& our static method.}
    \label{fig:reuse_comparison}
\end{figure*}

In the first part of the result analysis, we compare the reuse distance histograms generated dynamically by PARDA against our static approach.

Figure~\ref{fig:reuse_comparison} compares the reuse distance histograms for nested loop program snippets from the applications in Table~\ref{tab:test_applications}. The X-axis of each sub-figure represents the reuse distances, while the Y-axis shows the frequency of each reuse distance. Atax~\ref{fig:reuse_comparison}(a), Gemver~\ref{fig:reuse_comparison}(b), Mvt~\ref{fig:reuse_comparison}(c) and Trmm~\ref{fig:reuse_comparison}(d) shares a common trait; their maximum nested loop levels are two. That makes reuse distances and frequency counts predictable based on a second-degree equation with two variables. The total number of memory references of these programs are 106059, 2013609, 260709, and 16007009 respectively. Our static model predicts reuse profiles with 100\% accuracy for Atax, Gemver, and Trmm, while the accuracy is 92.37\% for Mvt, all compared to PARDA dynamic results.

In contrast, reuse profiles are harder to predict in cases with higher levels of nested loops, where some array references are reused after many other memory accesses irregularly. In Figure~\ref{fig:reuse_comparison}(e) (2mm), our static model failed to predict dilation correctly, resulting in a portion of missed reuse distance predictions and an overall accuracy of 86.74\% in predicting reuse distances out of 92402109 memory references. Similarly, Figure~\ref{fig:reuse_comparison}(f) (3mm) shows the model could not capture reuse distances of 106 and 2146, leading to a slightly reduced accuracy of 86.19\%. Still, our model predicts reuse profiles with reasonable accuracy overall.

\subsubsection{{Calculation Time}}
\begin{figure}[tb]
    \centering
    \includegraphics[width=0.49\textwidth]{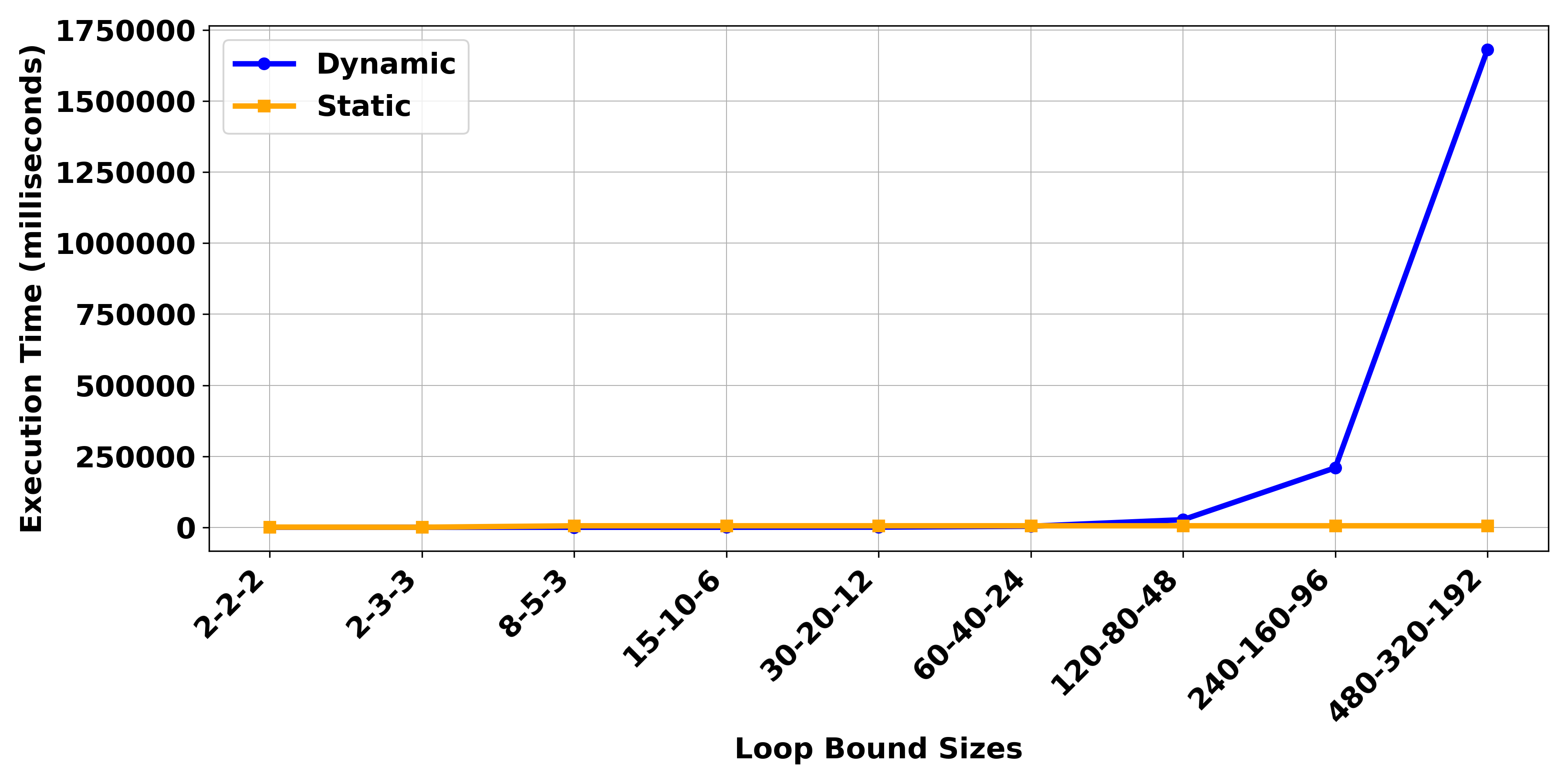}
    \caption{Reuse profile calculation time comparison for 2mm across increasing problem sizes (Loop bound: 2-2-2 to 480-320-192).}
    \label{fig:time_2mm}
\end{figure}

\begin{figure}[tb]
    \centering
    \includegraphics[width=0.49\textwidth]{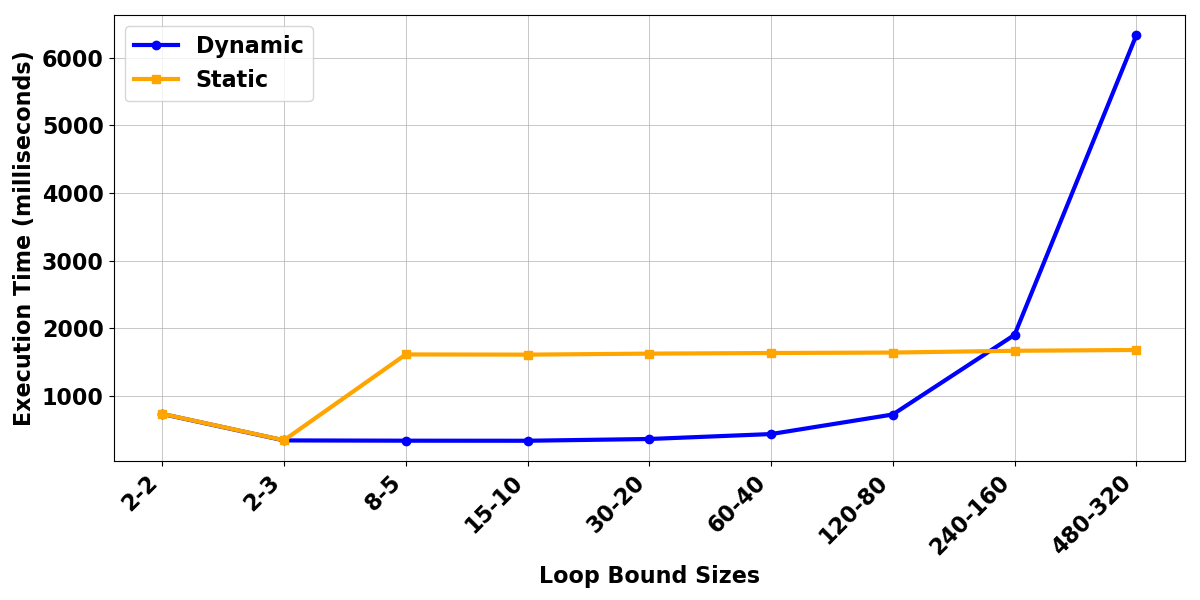}
    \caption{Reuse profile calculation time comparison for Atax calculating reuse distance histogram across increasing problem sizes (Loop bound: 2-2 to 480-320).}
    \label{fig:time_atax}
\end{figure}

The static predictor impressively computes the reuse profile almost in constant time across various problem sizes. As the problem size increases, the dynamic predictor's execution time grows significantly, especially for larger inputs, reflecting an exponential increase in computational complexity. Figures~\ref{fig:time_2mm} and~\ref{fig:time_atax} show that the time to profile the reuse distances for an application significantly explodes when problem size increases in the dynamic approach. In these figures, the X-axis represents the loop bounds growing from small to large, while the Y-axis shows the execution time in seconds. The blue line in the figures represents the execution time of PARDA's dynamic model across different loop bounds on the same problem. In contrast, the yellow line shows the running time of the same problem in our proposed static model.

Figure~\ref{fig:time_2mm} depicts the runtime comparison of the 2mm program with three nested loops. The smallest loop bounds are 2-2-2, meaning the outer, middle, and inner loops have two as the loop bound. In the same way, 480-320-192 represents the outer, middle, and innermost loop bounds, respectively. In smaller loop-bound scenarios, both methods perform similarly, but as the bounds increase, the dynamic model's execution time grows exponentially. On the other hand, the static approach maintains a consistent execution time. 

Similarly, Figure~\ref{fig:time_atax} compares the execution times for the Atax program with two nested loops. The execution time for the bound 8-5 increased slightly for the static model because, when the loop bounds exceed 3, it needs to calculate the reuse profiles for all the smaller loop bounds, such as 2-2, 2-3, 3-2, and 3-3, to set up the necessary variables for the mathematical equation. After that, for any larger loop bounds, the model relies solely on these smaller examples to predict the required bound.

The dynamic method's increased computation time is due to its need to run the entire program and collect memory traces for larger loop bounds. It waits to receive all dynamic memory references before calculating the profiles. In contrast, the static approach gathers a few reuse distance histograms from a few small loop bounds, and by making a mathematical correlation with the number at the smaller bounds, it extrapolates the reuse profile for larger bounds. Since the static approach uses the same equation to predict the reuse profile of the same program. However, with larger loop bounds or problem sizes, its execution time remains constant despite increasing the loop bounds.

\subsubsection{{Cache Hit Rate}}
\begin{figure}[tb]
    \centering
    \includegraphics[width=0.49\textwidth]{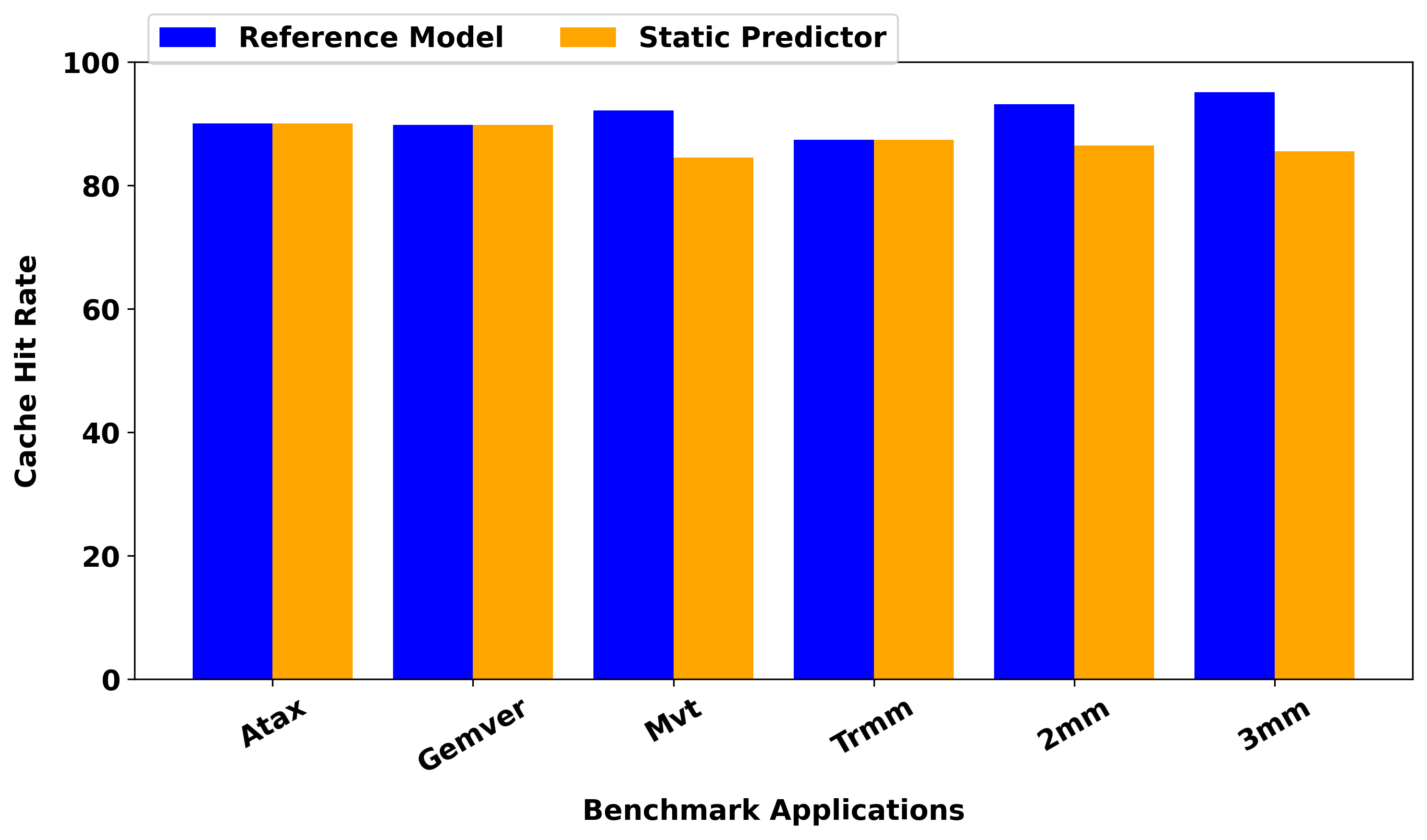}
    \caption{Cache Hit Rate comparison between the dynamic and static predictor.}
    \vspace{-.1cm}
    \label{fig:result-hit-rate}
\end{figure}

We calculate cache hit rates from the reuse profiles obtained using both methodologies and compare the results. Figure~\ref{fig:result-hit-rate} shows the hit rate result of both dynamic and static techniques in the given cache setup in section~\ref{subsubsec:cache-ref}. The X-axis lists the six benchmarks used in this study, while the Y-axis shows the corresponding cache hit rates. The blue bars represent the hit rates obtained from the dynamic method PARDA, while the yellow bars depict the results from our static method. 

Dynamic and static approaches for the Atax, Gemver, and Trmm benchmarks report cache hit rates of 90\%, 89.82\%, and 87.48\%, respectively. Our static approach precisely matches the dynamic reuse profiles, resulting in identical cache hit rates.

On the other hand, the static approach shows the cache hit rate for Mvt, 2mm, and 3mm as 84.48\%, 85.47\%, and 85.51\%, receptively, whereas the dynamic approach reports   92.11\%, 93.13\%, and 95.06\%, respectively. The discrepancy in hit rates is due to the static model's inaccuracies in estimating the reuse profile in the first place.  This gives slightly lower hit rates compared to the dynamic model. 
On average, our static approach accuracy is within $\sim95\%$ of the dynamic approach cache hit rate. This result is particularly impressive given the inherent challenges of static analysis in predicting changeable memory references. This high level of accuracy compared to the dynamic approach demonstrates the effectiveness of our static approach in characterizing the reuse profiles and cache hit rates of the benchmarks. Additionally, our approach is significantly faster than the dynamic reference approach.

\section{Limitations}
\label{sec:limitations}
There are a few limitations to our approach. Calculating from the innermost loop to the outer loops works correctly only for a few nested loops. When the loop depth increases, predicting the changeable reuse distance numbers becomes more challenging because the array references are being called more arbitrarily. Larger loop depth(s) would force us to use larger degree polynomials, which is not as easy as linear equations, but it is not impossible. Therefore, discrepancies will appear in the reuse profile compared to the dynamic approach for more loop nests. This is part of our future work.

This work does not address if-else branches within loops, which limits the model. It predicts large loop-bound reuse profiles from small cases without accounting for branch changes. When branch conditions are present inside the loop depending on the loop variable, static prediction of reuse profiles is also more challenging, as memory access patterns shift based on conditional directions. These are difficult to predict statically, but we could work on them using probabilities.

Additionally, this work is tailored for loop-based array programs and does not yet support large kernel programs.

\section{Conclusion}
\label{sec:conclusion}
Profiling applications is crucial for performance measurement, providing insights into program behavior, while reuse profiles and cache hit rates are essential metrics for evaluating memory systems efficiency. This paper introduces a static approach to predict reuse profiles of loop-based applications for larger loop bounds by analyzing reuse profiles of smaller loop bounds of the same application. Unlike dynamic approaches, our static method calculates reuse profiles without requiring running the benchmark and collecting full memory traces. Therefore, this approach is significantly faster in terms of execution time for estimating a program kernel's reuse profile and cache hit rate than the dynamic approach. Regarding prediction accuracy, our approach estimates reuse profiles that achieve cache hit rates averaging 95\% of those calculated by the dynamic approach (PARDA). Given that our static approach could miss capturing all reuse of non-uniform patterns of array references as the problem sizes grow, this model performs remarkably close to the dynamic approach accuracy while significantly reducing the time needed to provide the results.

\section*{Acknowledgment}
The authors would like to thank the anonymous reviewers. 
Triad National Security, LLC partially funded this research under subcontracts \#581326 and \#C4975. The Department of Energy (DOE) National Nuclear Security Administration (NNSA) under Contract DEAC52-06NA25396, and Laboratory Directed Research and Development (LDRD) program of Los Alamos National Laboratory under project YW0B00-PD24BARA also partially funded this research.

The opinions, findings, or conclusions expressed in this paper are solely the authors' and do not necessarily represent those of the DOE or the US Government.

\bibliographystyle{plainurl}
\bibliography{reference-arxiv}

\end{document}